\documentclass[twocolumn,aps, prl,showpacs]{revtex4}

\usepackage{graphics}
\usepackage{graphicx}
\usepackage{dcolumn}
\usepackage{bm}
\usepackage{amsmath,amssymb}
\usepackage{color}

\newcommand{\rev}[1]{{\color{red} #1}}

\begin{document}

\title{Coherent absorption of N00N states.}
\author{Thomas Roger$^{1}$, Sara Restuccia$^{2}$, Ashley Lyons$^{1}$,  Daniel Giovannini$^{2}$, Jacquiline Romero$^{2*}$,  John Jeffers$^3$, Miles Padgett$^{2}$,  Daniele Faccio$^{1}$}
\affiliation{$^{1}$School of Engineering and Physical Sciences, Heriot-Watt University, EH14 4AS Edinburgh, UK\\
$^{2}$School of Physics and Astronomy, University of Glasgow, Glasgow G12 8QQ, UK\\
$^{3}$Department of Physics, University of Strathclyde, Glasgow, G4 0NG, Scotland, UK\\
$^{*}$present address: School of Mathematics and Physics, University of Queensland, Brisbane, 4072, Australia}
\date{\today}
\begin{abstract}
{
Recent results in deeply subwavelength thickness films demonstrated coherent control and logical gate operations with both classical and single photon light sources. However, quantum  processing and devices typically involve more than one photon and non-trivial input quantum states.  Here we experimentally investigate two-photon N00N state coherent absorption in a multilayer graphene film. Depending on the N00N state input phase, it is possible to selectively choose between {\emph{single}} or {\emph{two}} photon absorption of the input state in the graphene film. These results demonstrate that coherent absorption in the quantum regime exhibits unique features opening up applications in multiphoton spectroscopy and imaging. 
}
\end{abstract}
\maketitle
\begin{figure}[t!]
\begin{centering}
\includegraphics[width = 0.5\textwidth]{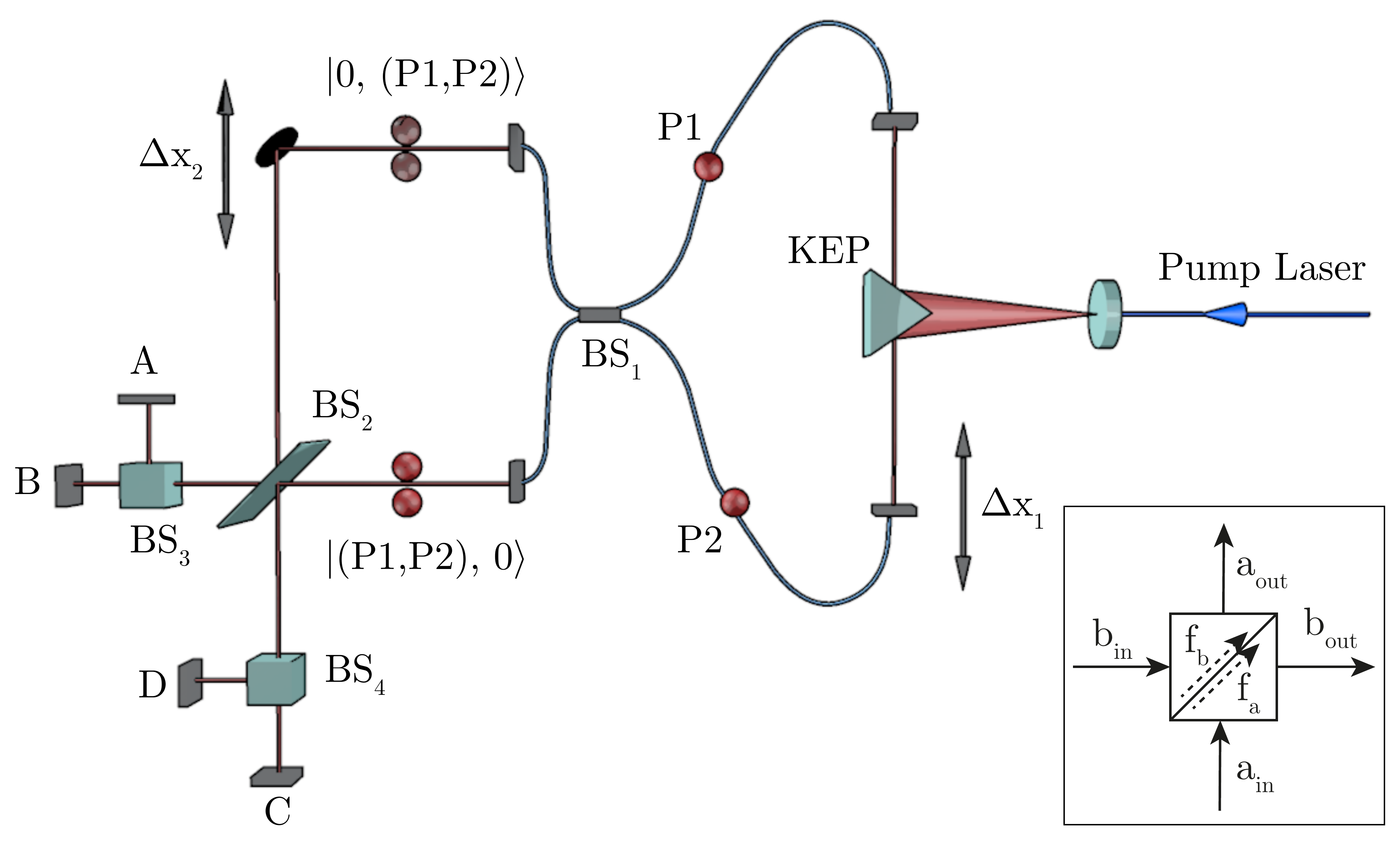}
\caption{\textbf{Experimental layout.} A frequency tripled Nd:YAG laser is used to pump a type 1 BBO crystal producing correlated biphotons. The biphotons (P1, P2) are separated by a knife-edge prism (KEP) and coupled to single-mode polarisation maintaining fibers. One of the fiber couplers is mounted on a translation stage in order to control the relative delay between the two photons. The photons are recombined on a 50:50 fiber beamsplitter BS$_1$ within their coherence length in order to produce bunched photon pairs which exit the beamsplitter in the quantum state $|2,0\rangle - e^{2i\phi}|0,2\rangle$. The phase $\phi$ of the quantum state is tuned by a mirror mounted to a piezo-electrically transduced stage. The quantum state is combined once again onto a second beamsplitter BS$_2$, which is either  a lossless 50:50 beamsplitter or  an absorptive graphene beamsplitter. The output states from the second beamsplitter are measured  with 4 single-photon avalanche dioides (SPADs), \emph{A}, \emph{B}, \emph{C} and \emph{D} after two further 50:50 beamsplitters, BS$_3$ and BS$_4$. The inset shows a schematic drawing of the output beamsplitter BS$_2$ with the input and output states, as referred to in the text.}
\label{fig:expt}
\end{centering}
\end{figure}
\begin{figure}[t!]
\begin{centering}
\includegraphics[width = 0.45\textwidth]{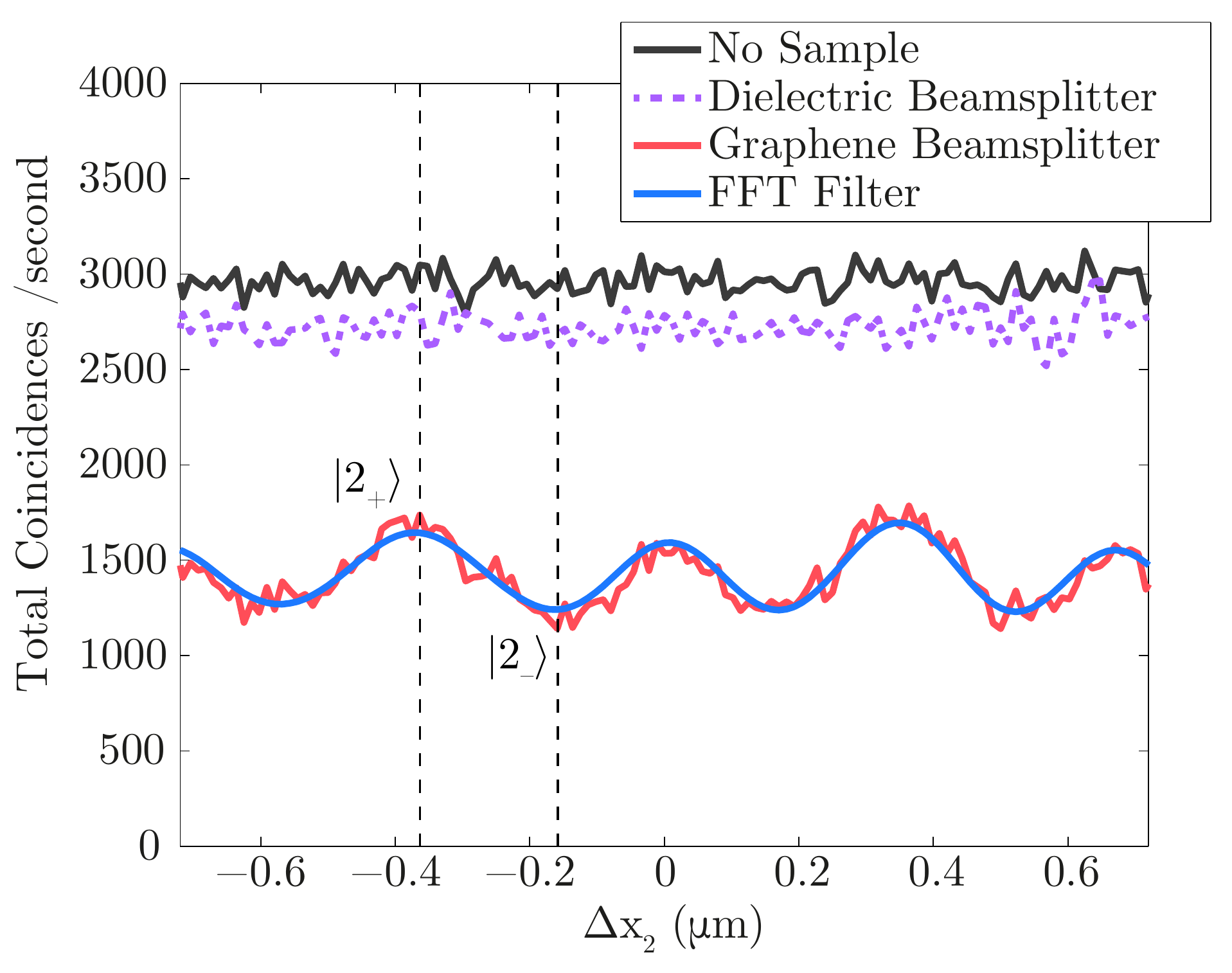}
\caption{\textbf{Experimentally measured total coincidence rates.}  Sum of coincidence counts for all six detection pairs as a function of interferometer mirror position measured for (i) no sample (black curve ), (ii) 50:50 lossless beamsplitter (purple curve) (iii) lossy graphene beamsplitter (red curve). The blue curve is a fit to the oscillation at frequency $k_2$. In (iii) the minima and maxima are related to a switching between input photon states $|2_+\rangle$ and $|2_-\rangle$, respectively, and therefore to  switching in the output states from single to two photon absorption. }
\label{fig:res3}
\end{centering}
\end{figure}

\emph{Introduction:} Coherent absorption is the process by which a partially absorbing material placed in a standing wave pattern can either perfectly transmit or absorb all of the incoming light. This process occurs as a result of the coherent interaction and of the relative phase relation of the two counter-propagating waves. Originally demonstrated in thick slabs of material where the process was likened to a ``time-reversed laser'' \cite{Wan, Chong}. Coherent absorption was recently extended to deeply subwavelength thickness (2D) materials that have 50\% absorption \cite{Fang, Dutta-Gupta, Zheludev}. Depending on the relative phase of the incident beams, it is possible to totally absorb (or transmit) light, thus providing a method to go beyond the theoretical limit of 50$\%$ absorption in a thin film \cite{Thongrattanasiri}, and therefore provide a route to obtain optical gates that rely on absorption \cite{zheludev}. This process relies heavily on the relations between reflection and transmission coefficients, $r$ and $t$, and absorption rate, $\alpha$, of the sample\cite{Jeffers}.  A special case is found for a sub-wavelength material exhibiting precisely 50$\%$ absorption, where complete absorption (or transmission) of the energy of the two beams is possible as a result of the relation between reflection and transmission coefficients being forced to $r = \pm{t}$ \cite{Thongrattanasiri}. This  has been shown experimentally with classical continuous wave sources \cite{Zheludev}, ultrashort pulses \cite{Rao}, and has also been extended to the nonlinear regime to achieve coherent control of nonlinear wave-mixing processes \cite{Rao2}. However, only a few studies have looked at the quantum nature of coherent absorption. \\
Recent experimental work has shown that a single photon also exhibits coherent perfect absorption \cite{roger,huang}, thus implying for example, deterministic absorption of the single photon itself.\\
More intriguing results are expected when non-trivial quantum states interact with a lossy beamsplitter. An N=2 N00N state, $(|N,0\rangle+e^{iN\phi}|0,N\rangle)/\sqrt{2}$ (obtained as the superposition of having N photons in one arm of an interferometer and none in the other) incident on a 50\% absorbing beamsplitter results in  states that can be composed of either single photons or a mixture of N-photon states (see Table 1) \cite{Jeffers}. Despite these theoretical predictions, this work is the first reported experiment investigating the coherent absorption of N00N states.   \\
Here we show for an input N=2 N00N state that controlling the relative phases of the two arms of an interferometer allows us to  either deterministically absorb only one photon or produce a mixed state which exhibits nonlinear absorption of two photons. \\
A Hong-Ou-Mandel (HOM) interferometer is used to create the N00N states that are then used as the input to a Mach-Zehnder interferometer (MZI).  The output beamsplitter of the MZI is a $\sim50\%$ absorbing multilayer graphene film. To explain the quantum coherent absorption process at the graphene beamsplitter, we first describe a theoretical model and then compare experimental results with N00N states incident on to a lossless and to a graphene beamsplitter. \\
\emph{Theoretical overview:}
The output states for a generic lossy beamsplitter (see inset to Fig.~\ref{fig:expt}) are related to the input states by, 
\begin{eqnarray}
\hat{a}_{\text{out}} & =& t\hat{a}_{\text{in}} +  r\hat{b}_{\text{in}} + \hat{f}_a\nonumber\\
\hat{b}_{\text{out}} &=& t\hat{b}_{\text{in}} +  r\hat{a}_{\text{in}} + \hat{f}_b
\end{eqnarray}
where $r$ is the reflection coefficient,  $t$ is the transmission coefficient, and $\hat{f}_{\text{a,b}}$ are the noise operators \cite{Barnett}.  The noise operators are necessary to preserve the commutators of the observable outputs for a generic beamsplitter which may exhibit loss or gain. In the case of a lossy beamsplitter we have $|t \pm r|< 1$ and the noise operators account for this loss.\\
\indent We consider a special case of lossy beamsplitter, with absorption  $A= 0.5$. This restriction forces the relation between reflected and transmitted waves to $r = \pm t=\pm0.5$ and therefore the phase difference between transmitted and reflected beams is restricted to $\theta = 0$ or $\theta= \pi$ \cite{Thongrattanasiri}. We consider the beamsplitter as shown in the inset to Fig.~\ref{fig:expt} with 50\% absorption, to which we input the superposition modes, $|2_+\rangle = (|2_a,0_b\rangle+|0_a,2_b\rangle)/\sqrt(2)$ or $|2_-\rangle = (|2_a,0_b\rangle-|0_a,2_b\rangle)/\sqrt(2)$. 
\begin{table}[t]
\caption{\textbf{Quantum states for a lossy/lossless beamsplitter.} Table showing the output photon states for various two-photon input states. }
\begin{center}
\begin{tabular}{l l c}
 \hline\hline $r$ \& $t$ phase & Input & Output \\ relation ($\theta$)  \vspace{0.1cm} \\  \hline 
$0$ or $\pi$ (50$\%$ loss) & $|2_+\rangle$ & $\frac{1}{2} |0_a, 0_b\rangle + \frac{1}{2}\left[\frac{1}{\sqrt{2}}(|1_a,1_b\rangle \pm |2_+\rangle)\right]$ \\
$0$ or $\pi$ & $|2_-\rangle$ & $\mp |1_\pm\rangle$\\ 
$\pi/2$ (lossless) & $|2_+\rangle$ & $|1_a,1_b\rangle$\\ 
$\pi/2$ & $|2_-\rangle$ & $\mp|2_{\pm}\rangle$ \\ \hline
\end{tabular}
\end{center}
\label{tab:1}
\end{table}
The output states for a 50\%-loss beamsplitter have been derived in Ref.~ \cite{Jeffers}  and are summarised in Table \ref{tab:1}. In row 1 we see that for a $|2_+\rangle$ input state either both photons survive interaction with the sample or both photons are absorbed (by the noise operator, $\hat{f}$). Remarkably, this implies that the sample exhibits pure two-photon absorption, a process that typically occurs only in the nonlinear (i.e. high beam intensity) regime but here arises as a result of coherent absorption acting upon a specific quantum photon state. On the other hand, for a $|2_-\rangle$ input state, one photon of each pair is always absorbed,  hence the output is a superposition of one-photon states that \emph{must} contain only one photon.\\
For a lossy splitter with $r \neq t$, the phase relations deviate from the ideal case described above. This has the effect of mixing the output modes, which means that there will be zero-, one- and two-photon components at the output, with ratios determined non-trivially by the relevant combinations of the reflection and transmission coefficients. In real experiments, this implies that we will effectively reduce the purity of the single  and two-photon absorption processes.\\
To complete the picture we also include in Table I (third and fourth rows) the results for the lossless case where $t = ir$ i.e. $\phi = \pi/2$ (where $\phi$ is the input phase), as demonstrated in a cascaded Mach-Zehnder interferometer \cite{itoh}. We see that depending on the input phase, the bi-photons either bunch together or split apart, therefore conserving total photon number, as should be expected when the noise operators are removed.\\
 \indent \emph{Experiments:} We use a frequency tripled Nd:YAG laser producing $>$15 ps pulses centred at $\lambda = 355$ nm at a repetition rate of 120 MHz (downsampled to 60 MHz),  to pump a type 1 $\beta$-barium-borate ($\beta$-BBO) crystal and create correlated single photon pairs  with a central wavelength $\lambda = 710$ nm. The photon pairs are separated in the far field and collected with polarisation maintainining fibers. These fibers are combined onto a lossless 50:50 beamsplitter BS$_1$ (the first BS in our Mach-Zehnder interferometer, see Fig.~\ref{fig:expt}).  A controllable delay on one of the input arms to BS$_1$ controls the path length difference between the two  arms, which is optimised so that the photons bunch as shown in the famous experiment of Hong, Ou and Mandel \cite{hong}. The bunched biphotons serve as the input to the second beamsplitter BS$_2$, which corresponds to the device shown in the inset to  Fig.~\ref{fig:expt}. BS$_2$ is either (i) an absorptive graphene film or (ii) a lossless 50:50 beamsplitter. The multilayer graphene sample is grown by chemical vapour deposition from Graphene Platform with approximately 40 layers of graphene on a substrate of fused silica. This provides an absorption of $\sim 61 \%$ and reflection and transmission coefficients of $r = 0.30$ and $t = 0.55$ respectively.
 \begin{figure}[t!]
\begin{centering}
\includegraphics[width = 0.5\textwidth]{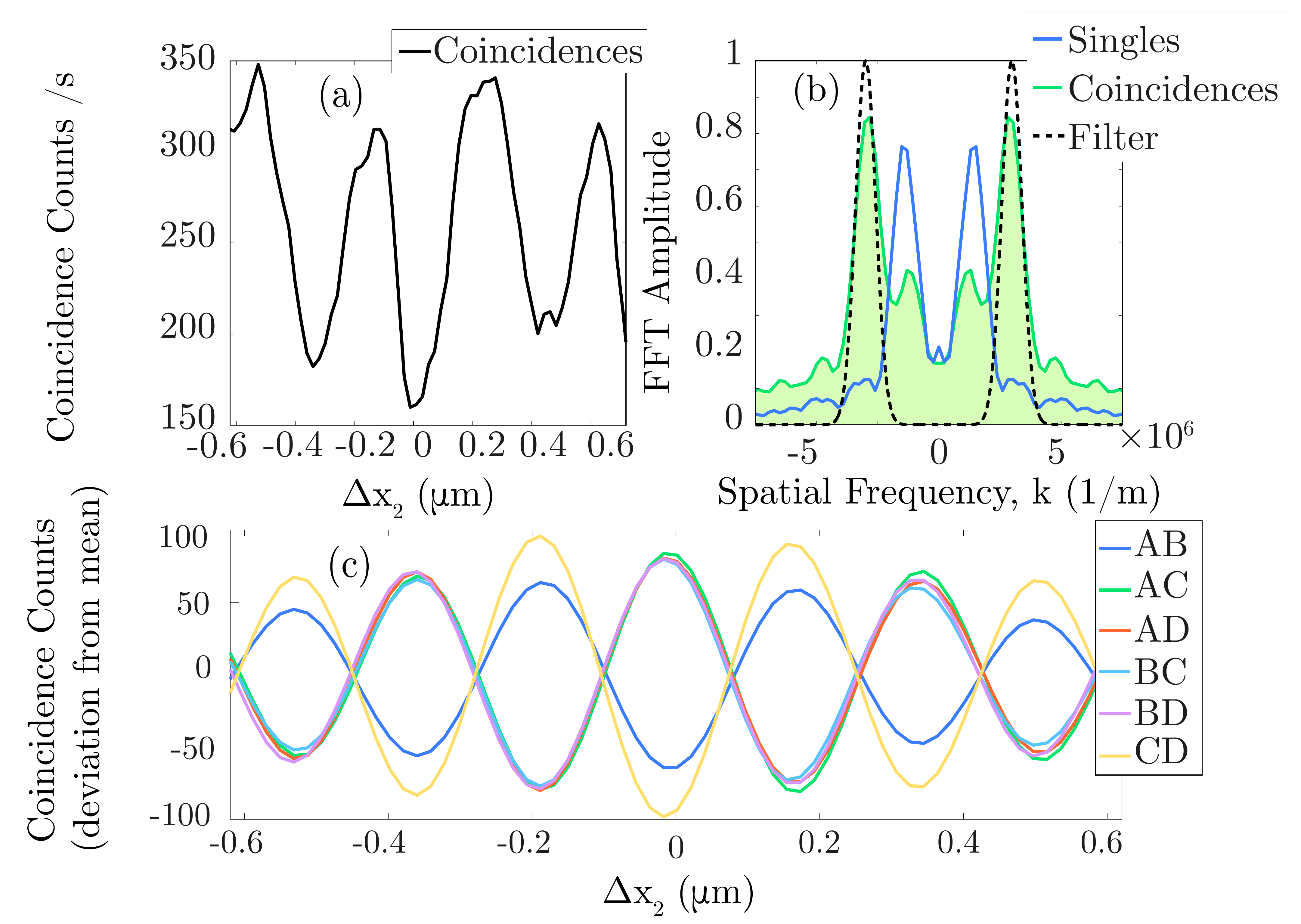}
\caption{\textbf{Lossless 50:50 beamsplitter.} (a) Raw experimental data showing the coincidence counts per second measured between detectors \emph{A} and \emph{B} as a function of interferometer mirror position. (b) Blue curve, Fourier transform of the {\emph{single}} detector counts showing an oscillation peak at $k_1 = 1/\lambda$. Green shaded area curve, Fourier transform of the {\emph{coincidence}} counts, showing a peak at frequency $k_2 = 2/\lambda$. Dashed black curve - gaussian fit to the peak at $k_2$ used to filter the raw data. (c) Fourier-filtered data for all six detection pairs as a function of mirror position. Same side detection pairs (\emph{AB} \& \emph{CD}) are $\pi$ out-of-phase with respect to opposing side detector pairs (e.g. \emph{BC} \& \emph{AD}).}
\label{fig:res1}
\end{centering}
\end{figure}
\begin{figure}[t!]
\begin{centering}
\includegraphics[width = 0.5\textwidth]{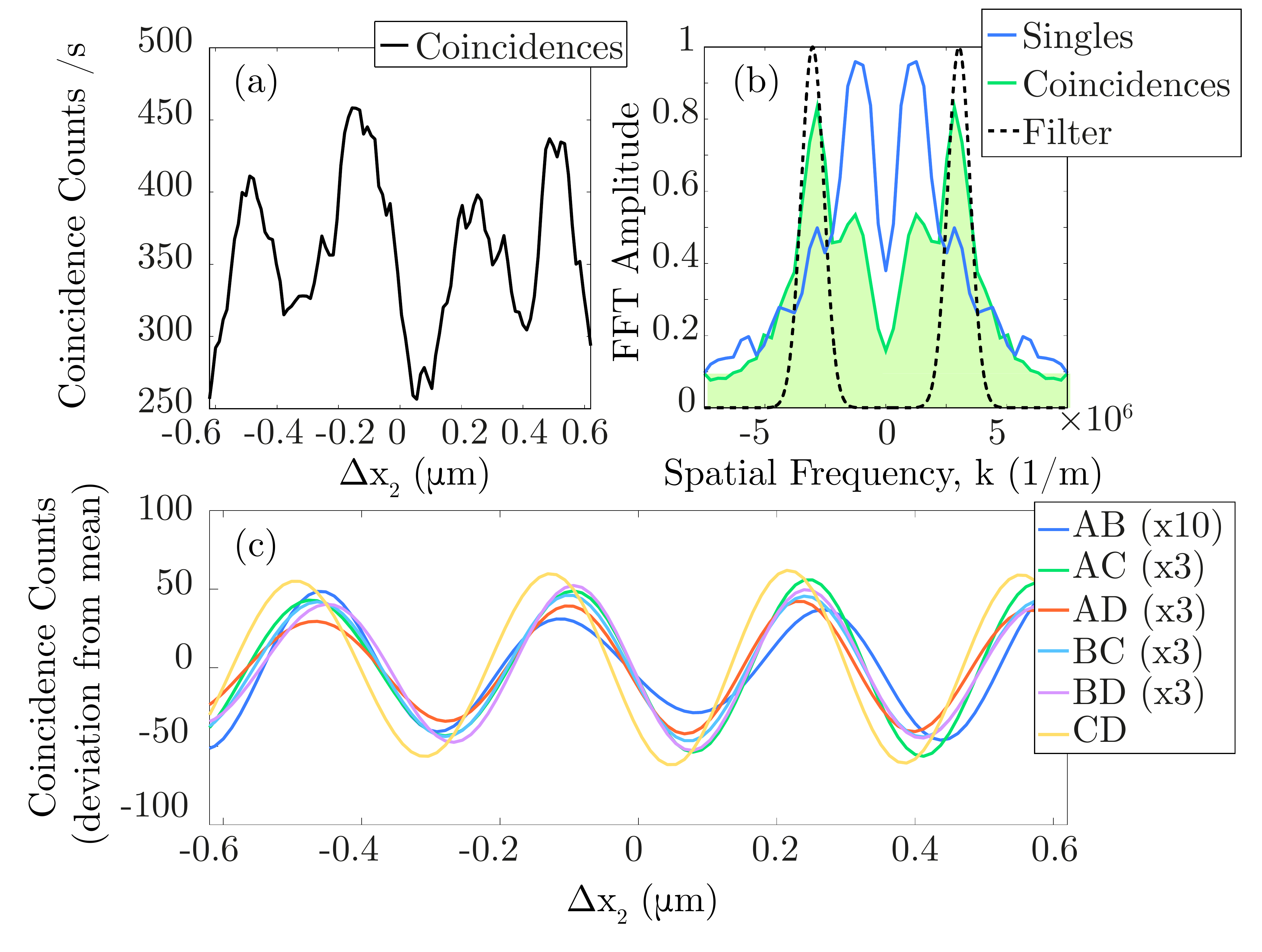}
\caption{\textbf{Experimental data from a lossy beamsplitter.} (a) Raw  data showing the coincidence counts per second measured between detectors \emph{A} and \emph{B} as a function of interferometer mirror position. (b) Blue curve, Fourier transform of the {\emph{single}} detector counts showing an oscillation at $k_1 = 1/\lambda$. Green shaded curve, Fourier transform of the {\emph{coincidence}} counts showing a peak at frequency $k_2 = 2/\lambda$. Dashed black curve, gaussian fit to the peak at $k_2$ used to filter the raw data. (c) Fourier-filtered data for all six detection pairs as a function of mirror position. All 6 coincidence traces are in phase, suggesting a coherent absorption effect. The slight phase shift seen here is due to the asymmetry of the air-graphene-substrate beamsplitter.}
\label{fig:res2}
\end{centering}
\end{figure}
A piezo-electric stage on one of the interferometer mirrors controls the phase of the N00N states incident on BS$_2$. The photons in each of the output ports of BS$_2$ are split once again in the two 50:50 beamsplitters, BS$_3$ and BS$_4$, resulting in four detection ports which we label \emph{A}, \emph{B}, \emph{C} and \emph{D}. The output photons are detected with single photon avalanche detectors (SPADs) connected to a PicoQuant Hydra Harp 400 detection card allowing time-tagged events to be recorded. Measurements are performed in time-tagged time-resolved mode  such that the coincident photon events can subsequently be analysed in multiple configurations (\emph{AB}, \emph{AC}, ..., \emph{CD}). We set the coincidence time window to $\Delta\tau = 25$ ns. A complete set of coincidence measurements are taken as a function of the interferometer phase across all four detection ports.\\
\indent \emph{Results:}
Figures~\ref{fig:res1} and \ref{fig:res2} show the results for the cases of the lossless and graphene (lossy) beamsplitter, respectively. Both figures show examples of raw coincidence count data (a), and the Fourier transforms of the data (b). The latter highlights the oscillation in the coincidences at twice the frequency of the single detector counts, as expected with input N00N states.  We then isolate the N00N state contribution by multiplying the data by a Gaussian-shaped filter [black dashed line in (b)] and inverse Fourier transforming so as to obtain the curves in (c). The curves in Figures~\ref{fig:res1}(c) and \ref{fig:res2}(c) show the N00N coincidence counts across all detector pairings, as indicated in the figures. We underline the main difference between the lossless and graphene beamsplitter measurements: whilst in the lossless case the coincidence counts corresponding to detectors located on opposite sides of the beamsplitter (\emph{AC}, \emph{AD}, \emph{BC}, \emph{BD}) are out of phase with those from detectors on the same side (\emph{AB} and \emph{CD}), in the graphene case all pairings oscillate in phase. This is in agreement with the theoretical predictions summarised in Table \ref{tab:1} and is a clear indication of the action of loss on the input N00N states. 
Indeed, for an input $|2_+\rangle$ state we predict and observe an equal probability of coincidences being measured between same-side \emph{and} opposing-side detection ports. Therefore all detector pairs see a maximum. For the $|2_-\rangle$ input state,  a single photon is absorbed so that it is not possible to measure a coincidence between \emph{any} of the detection ports, and as a result we see a minimum for all detector pairs. This is the key observation that lies at the heart of the very peculiar features of coherent absorption of N00N states. \\
\indent  An alternative way of viewing this data is provided in Fig.~\ref{fig:res3} that shows the sum of the coincidences measured across all 6 detection pairs ($2AB+AC+AD+BC+BD+2CD$) as a function of the interferometer mirror position $\Delta\text{x}_2$ for three cases. We note that when adding the coincidence counts together,  it is necessary to multiply the same side coincidence counts $AB$ and $CD$ by a factor of 2 in order to account for the fact that 50\% of the time, two photons incident on the output beamsplitter (BS$_3$ or BS$_4$) will bunch together  and therefore be recorded as a single click at a single detector. For the case where there is no beamsplitter or a lossless beamsplitter (solid black and dashed purple curve in Fig.~\ref{fig:res3}, respectively), there is no change in the coincidence counts with interferometer mirror position, as expected.   What we measure is merely a re-distribution of the photons within the system that is controlled by the relative lengths of the interferometer arms. The slight reduction in total coincidence rate with respect to the `no beam-splitter' case is accounted for by a small amount of loss inherent in the beamsplitter. Conversely, in the case of the graphene beamsplitter, we see an oscillation of the total coincidence rate as a function of the interferometer phase. This is due to the change in the total energy within the system as predicted also by the quantum states  in Table \ref{tab:1}. We note that the modulation depth  is reduced with respect to the ideal case, which we ascribe to the non-ideal ($>50\%$) absorption of our beamsplitter. However, the main traits summarised in  Table \ref{tab:1} are still clearly visible: as the input states are  changed from $|2_-\rangle$ to $|2_+\rangle$,  we transition from an increased likelihood of single-poton to two-photon absorption, respectively.
\indent \emph{Conclusions:}  We show coherent absorption of N00N states on a \emph{lossy} beamsplitter exhibits very distinct features compared to single photon or classical beam coherent absorption. If the lossy beamsplitter absorbs 50\%, then single photon states can be either completely absorbed or transmitted. Conversely, with input two-photon N00N states by varying the input states we can selectively control whether only a single photon is absorbed or two photons are absorbed. The fact that a two-photon state may exhibit full two-photon absorption alone (i.e. with no single-photon absorption) is a purely quantum effect and is in itself rather remarkable. In a certain sense our work is reminiscent of two-photon absorption processes in nonlinear optics yet here, all processes are inherently linear. 
Loss is usually highly detrimental to quantum states but may also be envisaged as a resource for quantum photonics and applications involving for example N00N states in quantum metrology  \cite{bell,rozema,silberberg,jacobsen} and the ability to control single versus two-photon absorption may be used to selectively isolate single and two-photon processes in individual molecules or quantum dots \cite{molecules}.
\section*{Acknowledgements}
D. F. acknowledges financial support from the European Research Council under the European Unions Seventh Framework Programme (FP/2007-2013)/ERC GA 306559 and EPSRC (UK, Grant EP/J00443X/1). S. R. acknowledges support from ERC Advance grant (TWISTS).

\end{document}